\documentclass[11pt, a4paper]{article}
\usepackage[a4paper, top=1cm, bottom=2.5cm, left=3cm, right=3cm]{geometry}

\usepackage{iftex}
\ifLuaTeX
  \usepackage{fontspec}
  \usepackage[english, bidi=basic, provide=*]{babel}
  \babelprovide[import, onchar=ids fonts]{english}
  \babelfont{rm}{Noto Sans}
  \babelfont{sf}{Noto Sans}
  
\else
  \usepackage[T1]{fontenc}
  \usepackage[utf8]{inputenc}
  \usepackage{helvet}
  
  \usepackage[english]{babel}
\fi

\usepackage{amsmath}
\usepackage{xcolor} 
\usepackage{array}  
\usepackage{booktabs}
\usepackage{caption}
\usepackage{float}
\usepackage{tabularx}
\usepackage{enumitem}
\usepackage{parskip}

\setlist{nosep} 

\usepackage{tikz}
\usetikzlibrary{shapes.geometric, arrows.meta, positioning, calc, shadows}
\usepackage{pgfplots}
\pgfplotsset{compat=1.18}

\usepackage[colorlinks=true, linkcolor=blue, citecolor=blue, urlcolor=blue]{hyperref}

\title{\textbf{Agentic Compilation: Mitigating the LLM Rerun Crisis for Minimized-Inference-Cost Web Automation}}
\author{Jagadeesh Chundru \\ 
\small \href{mailto:jagadeesh@selfotix.com}{\textcolor{black}{\texttt{jagadeesh@selfotix.com}}} \\
\small Selfotix}
\date{}

\begin{document}

\maketitle

\section*{Abstract}
LLM-driven web agents operating through continuous inference loops---repeatedly querying a model to evaluate browser state and select actions---exhibit a fundamental scalability constraint for repetitive tasks. We characterize this as the Rerun Crisis: the linear growth of token expenditure and API latency relative to execution frequency. For a 5-step workflow over 500 iterations, a continuous agent incurs approximately \$150.00 in inference costs; even with aggressive caching, this remains near \$15.00. 

We propose a Compile-and-Execute architecture that decouples LLM reasoning from browser execution, reducing per-workflow inference cost to under \$0.10. A one-shot LLM invocation processes a token-efficient semantic representation from a DOM Sanitization Module (DSM) and emits a deterministic JSON workflow blueprint. A lightweight runtime then drives the browser without further model queries. 

We formalize this cost reduction from $O(M \times N)$ to amortized $O(1)$ inference scaling, where $M$ is the number of reruns and $N$ is the sequential actions. Empirical evaluation across data extraction, form filling, and fingerprinting tasks yields zero-shot compilation success rates of 80--94\%. Crucially, the modularity of the JSON intermediate representation allows minimal Human-in-the-Loop (HITL) patching to elevate execution reliability to near-100\%. At per-compilation costs between \$0.002 and \$0.092 across five frontier models, these results establish deterministic compilation as a paradigm enabling economically viable automation at scales previously infeasible under continuous architectures.

\section{Introduction}

The automation of web-based workflows has evolved from brittle, hand-crafted scripts toward adaptive AI-driven agents capable of interpreting dynamic interfaces and responding to unforeseen UI states. Contemporary approaches frequently adopt continuous-loop architectures exemplified by frameworks such as ReAct (Yao et al., 2022) and Chain-of-Thought (Wei et al., 2022) in which an LLM is invoked at each discrete step to observe the current browser state and emit the next action. While this paradigm affords considerable flexibility, it incurs a computational cost that scales multiplicatively with both task complexity and execution frequency.

This paper identifies, formalizes, and addresses a specific failure mode of continuous-loop agents that we term the Rerun Crisis. The Rerun Crisis arises when a fixed workflow---one whose logical structure does not change between executions---is repeated across many iterations, each of which independently queries the LLM to re-derive an already-known action sequence. This redundancy transforms a constant-cost planning problem into a linearly scaling inference expenditure, rendering continuous agents economically untenable for enterprise-scale automation. 

To resolve this, we introduce One-Shot Agentic Compilation, a systems-level architectural shift that confines LLM involvement to a single initialization phase. The model receives a semantically pruned representation of the target web page and the user's intent, and emits a structured JSON workflow blueprint as an intermediate representation (IR). Execution is subsequently delegated to a deterministic execution engine, eliminating all further model queries. This architecture mirrors the historical shift from interpreted scripting to compiled binaries, or dynamic execution to static query planning in databases. 

The contributions of this work are as follows: 
\begin{itemize}
    \item We formalize the Rerun Crisis as a cost-scaling problem, deriving its $O(M \times N)$ inference complexity and quantifying its financial impact under representative enterprise workloads. 
    \item We present the DOM Sanitization Module (DSM), a token-reduction preprocessing stage that compresses raw HTML by up to 85\% through noise removal, visibility filtering, and attribute cleansing. 
    \item We propose a zero-shot compilation protocol that forces the LLM to perform complex spatial and semantic reasoning ahead of time---including robust selector engineering, zero-shot pagination inference, and structural loop deduction---generating a deterministic JSON blueprint from a single inference call. 
    \item We report empirical benchmarks across five frontier models and three task modalities, establishing the practical viability of the proposed architecture. 
\end{itemize}

\noindent\fbox{
    \parbox{\dimexpr\textwidth-2\fboxsep-2\fboxrule\relax}{
        \textbf{Definition (Rerun Crisis):}\\
        The Rerun Crisis refers to the redundant recomputation of deterministic workflows in continuous-loop agents, leading to linear growth in inference cost with respect to execution frequency.
    }
}

\subsection{The Rerun Crisis}

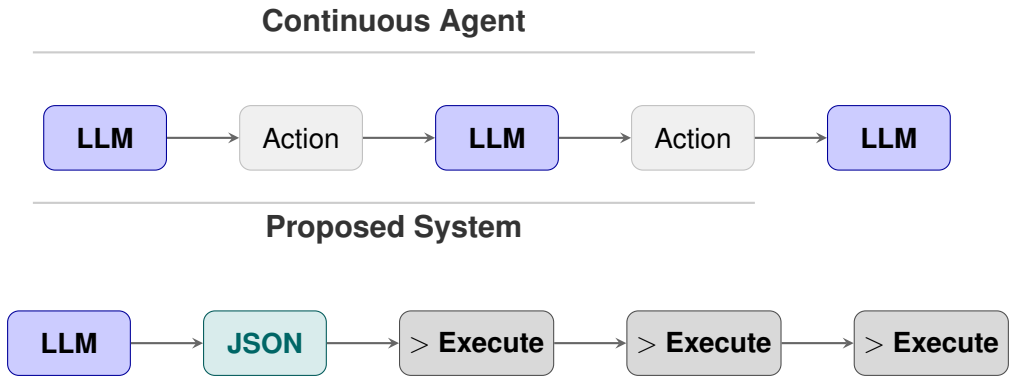
\begin{figure}[H]
\centering
\begin{tikzpicture}[
    node distance=1.3cm and 1cm,
    >=stealth,
    scale=0.95,
    transform shape
]


\tikzstyle{llm} = [
    rectangle, rounded corners=0.15cm,
    minimum width=1.7cm, minimum height=0.9cm,
    text centered,
    draw=blue!60!black,
    fill=blue!20,
    text=black,
    font=\bfseries\sffamily
]

\tikzstyle{action} = [
    rectangle, rounded corners=0.15cm,
    minimum width=1.7cm, minimum height=0.9cm,
    text centered,
    draw=gray!50,
    fill=gray!12,
    text=black,
    font=\sffamily
]

\tikzstyle{json} = [
    rectangle, rounded corners=0.15cm,
    minimum width=1.7cm, minimum height=0.9cm,
    text centered,
    draw=teal!70!black,
    fill=teal!15,
    text=teal!80!black,
    font=\bfseries\sffamily
]

\tikzstyle{exec} = [
    rectangle, rounded corners=0.15cm,
    minimum width=1.7cm, minimum height=0.9cm,
    text centered,
    draw=black!70,
    fill=black!15,
    text=black,
    font=\bfseries\sffamily
]


\node[font=\bfseries\large\sffamily, text=black!80] (title1) {Continuous Agent};
\draw[black!20, thick] (-5,-0.4) -- (5,-0.4);


\node[llm, below=0.8cm of title1, xshift=-4cm] (l1) {LLM};
\node[action, right=of l1] (a1) {Action};
\node[llm, right=of a1] (l2) {LLM};
\node[action, right=of l2] (a2) {Action};
\node[llm, right=of a2] (l3) {LLM};

\draw[->, thick, draw=black!60] (l1) -- (a1);
\draw[->, thick, draw=black!60] (a1) -- (l2);
\draw[->, thick, draw=black!60] (l2) -- (a2);
\draw[->, thick, draw=black!60] (a2) -- (l3);


\node[font=\bfseries\large\sffamily, text=black!80, below=2.2cm of title1] (title2) {Proposed System};
\draw[black!20, thick] (-5,-2.5) -- (5,-2.5);

\node[llm, below=0.8cm of title2, xshift=-4.5cm] (pl1) {LLM};
\node[json, right=of pl1] (pj1) {JSON};
\node[exec, right=of pj1] (pe1) {$>$ Execute};
\node[exec, right=of pe1] (pe2) {$>$ Execute};
\node[exec, right=of pe2] (pe3) {$>$ Execute};

\draw[->, thick, draw=black!60] (pl1) -- (pj1);
\draw[->, thick, draw=black!60] (pj1) -- (pe1);
\draw[->, thick, draw=black!60] (pe1) -- (pe2);
\draw[->, thick, draw=black!60] (pe2) -- (pe3);

\end{tikzpicture}
\caption{Architectural comparison between continuous-loop reasoning and the proposed one-shot Compile-and-Execute paradigm.}
\label{fig:comparison}
\end{figure}

Let $N$ denote the number of sequential browser actions comprising a single workflow, and $M$ the number of times that workflow is executed. In a continuous-loop agent, each of the $N$ steps in each of the $M$ executions requires an independent LLM inference call over a context that includes the current DOM state. If $S_i$ denotes the token size of the DOM snapshot at step $i$, and $C_t$ the per-token inference cost, the total expenditure is: 

\begin{equation}
\text{Cost}_{cont} = M \times \sum_{i=1}^{N} [S_i \times C_t]
\end{equation}

This expression scales as $O(M \times N)$, meaning that doubling either the workflow length or the number of executions doubles the inference cost. For high-volume enterprise applications---such as extracting structured records from thousands of paginated listings, or submitting bulk form payloads---this scaling rapidly becomes prohibitive. The problem is not one of agent intelligence or task difficulty; it is a structural inefficiency arising from the reapplication of probabilistic reasoning to deterministic, previously solved problems. 

\subsection{Empirical Motivation from Production Deployments}

The Rerun Crisis is the unspoken bottleneck of modern AI engineering. While model capabilities and context windows have expanded massively, applying continuous reasoning to static enterprise tasks creates a financial paradox: workflows succeed in prototype but fail in production. Industry discourse is saturated with reports of engineering teams abandoning continuous-loop frameworks not due to logical failures, but due to exhausted billing quotas, unpredictable token scaling, and severe latency bottlenecks. 

Community benchmarks of prominent open-source continuous agents record per-task inference costs ranging from \$1.00 to \$3.20 for routine workflows when routed through frontier models. In high-volume enterprise pipelines, this translates to thousands of dollars for tasks that traditional RPA executed for pennies. The industry has largely mischaracterized this as a transient "pricing issue," waiting for API costs to fall. We argue it is fundamentally an architectural flaw. 

This work is not intended to introduce a new learning algorithm, but rather to propose a systems-level architectural shift that reframes LLMs as compilers rather than continuous decision-makers. By giving the Rerun Crisis a formal definition, we aim to shift the field's focus from brute-forcing prompt optimization toward designing structurally efficient execution paradigms.

\section{Related Work}

\subsection{Continuous-Loop Agentic Frameworks}

The dominant paradigm in agentic web automation research builds upon iterative reasoning frameworks. ReAct (Yao et al., 2022) interleaves language model reasoning with environment interaction, producing flexible agents capable of handling diverse tasks. Chain-of-Thought prompting (Wei et al., 2022) extends this by eliciting intermediate reasoning steps, improving performance on complex multi-hop tasks. Subsequent work has applied these principles to browser-based agents using both text-based DOM representations and visual inputs via Vision-Language Models (VLMs). While highly adaptive, these architectures impose the $O(M \times N)$ inference cost structure described in Section 1.1. Production-oriented mitigations including DOM diffing, prompt compression, and partial state caching can reduce per-step costs but do not alter the fundamental scaling behavior. Even at a theoretically optimistic 90\% caching efficiency, total inference costs remain proportional to the product of task length and execution count. Recent rigorous evaluations have further identified that continuous-loop agents suffer from context degradation as trajectory lengths increase, and that reported benchmark gains may not transfer reliably to real-world environments (Xue et al., 2025). 

\subsection{LLM-Assisted Script Generation and Workflow Memory}

An alternative approach employs LLMs in a one-shot capacity to generate traditional automation scripts typically in Python or via browser automation libraries such as Playwright or Selenium (Chen et al., 2021). This strategy successfully eliminates per-rerun inference costs, as the generated script executes without further model involvement. However, it introduces substantial deployment friction: generated scripts require language runtimes, dependency management, and execution environment configuration, limiting accessibility in enterprise contexts where infrastructure standardization is constrained. 

Recent approaches such as Agent Workflow Memory (AWM) (Zheng et al., 2024) attempt to optimize repetitive tasks by inducing commonly reused routines from past experiences and supplying them as context. While AWM reduces the number of exploratory steps ($N$) by utilizing memory, it still requires the LLM to actively process and execute the steps at runtime, remaining trapped in the $O(M \times N)$ continuous-loop paradigm. In contrast, our Compile-and-Execute architecture emits a declarative JSON workflow schema evaluated by a purpose-built runtime engine. This eliminates the LLM entirely during execution, achieving true amortized $O(1)$ scaling while avoiding the deployment overhead associated with arbitrary code execution.

\subsection{DOM Pruning and Context Reduction}

A significant obstacle to LLM-based web interaction is the size of raw DOM representations, which frequently range from 10,000 to over 100,000 tokens and exceed the effective context windows of current models. Prior work has proposed programmatic filtering strategies that selectively retain semantically relevant DOM elements prior to model invocation (Zhang et al., 2025). Our DOM Sanitization Module (DSM) extends this line of work by applying structured pruning specifically to support downstream compilation into an executable schema, with an emphasis on preserving selector robustness and pagination structure over general-purpose readability. 

\section{Methodology and System Architecture}

The proposed architecture is organized as a three-stage pipeline: (1) context optimization via DOM sanitization, (2) one-shot LLM compilation, and (3) supervised deterministic execution. These stages correspond to a clean separation of concerns: the LLM handles reasoning and intent translation; the DSM handles context preparation; and the execution engine handles deterministic browser control. A Human-in-the-Loop (HITL) verification gate is interposed between compilation and execution to support auditability and safety. 

To formalize the compile-and-execute paradigm, we explicitly map the proposed architecture to standard compiler theory: 
\begin{itemize}
    \item Natural language user intent serves as the Source Code. 
    \item The one-shot LLM functions as the Compiler, utilizing its spatial reasoning to parse the intent. 
    \item The resulting deterministic JSON Blueprint acts as the Bytecode or Intermediate Representation (IR). 
    \item The lightweight Execution Engine operates as the Runtime Environment. 
\end{itemize}

This explicit separation of compilation from runtime execution is what enables the framework's strict amortized $O(1)$ inference cost-scaling.

\begin{figure}[H]
\centering
\begin{tikzpicture}[
    node distance=1.2cm and 0.8cm, 
    >=stealth, 
    scale=0.85, 
    transform shape
]


\tikzstyle{basebox} = [
    rectangle, rounded corners=0.18cm,
    draw=gray!60!blue,
    fill=gray!8,
    text=black,
    align=center,
    minimum height=1.2cm,
    minimum width=3.6cm,
    font=\sffamily,
    drop shadow={opacity=0.12}
]

\tikzstyle{intelligencebox} = [
    rectangle, rounded corners=0.22cm,
    draw=violet!90!black,
    fill=violet!75!black,
    text=white,
    align=center,
    minimum height=1.3cm,
    minimum width=4.4cm,
    font=\bfseries\sffamily,
    drop shadow={opacity=0.2}
]

\tikzstyle{outputbox} = [
    rectangle, rounded corners=0.18cm,
    draw=green!60!black,
    fill=green!8,
    text=green!50!black,
    align=center,
    minimum height=1.2cm,
    minimum width=3.6cm,
    font=\bfseries\sffamily,
    drop shadow={opacity=0.12}
]

\tikzstyle{decisionbox} = [
    rectangle, rounded corners=0.18cm,
    draw=orange!85!black,
    fill=orange!12,
    text=orange!90!black,
    align=center,
    minimum height=1.2cm,
    minimum width=3.6cm,
    font=\bfseries\sffamily,
    drop shadow={opacity=0.12}
]

\tikzstyle{browserbox} = [
    rectangle, rounded corners=0.22cm,
    draw=black!85,
    fill=black!80,
    text=white,
    double,
    double distance=1pt,
    align=center,
    minimum height=1.2cm,
    minimum width=3.6cm,
    font=\bfseries\sffamily,
    drop shadow={opacity=0.25}
]


\node[basebox] (target) at (0,0) {\textbf{Target Web Page}\\Raw HTML};

\node[basebox] (dsm) at (5,0) {
\textbf{DOM Sanitization}\\
\textbf{Module}\\
\scriptsize Prunes Noise, Extracts Semantics
};

\node[basebox] (skeleton) at (10,0) {\textbf{Semantic Skeleton}};

\node[intelligencebox] (llm) at (5,-2.6) {
One-Shot LLM Compiler\\
\scriptsize Generates Deterministic Script
};

\node[basebox] (browser) at (13.5,-2.6) {\textbf{Browser}};

\node[basebox] (intent) at (0,-5.2) {\textbf{User Intent / Prompt}};

\node[outputbox] (json) at (5,-5.2) {JSON Blueprint};

\node[decisionbox] (hitl) at (10,-5.2) {
\textbf{HITL Verification Gate}\\
\scriptsize Human Approval
};

\node[browserbox] (action) at (13.5,-7.8) {
Browser Action\\
\scriptsize Clicks/Inputs\\
\scriptsize Zero AI Inference
};


\draw[->, thick, draw=black!60] (target) -- (dsm);
\draw[->, thick, draw=black!60] (dsm) -- (skeleton);
\draw[->, thick, draw=black!60] (skeleton) -- (llm);
\draw[->, thick, draw=black!60] (dsm) -- (llm);
\draw[->, thick, draw=black!60] (intent) -- (llm);

\draw[->, thick, draw=black!60] (llm) -- (json);
\draw[->, thick, draw=black!60] (json) -- (hitl);
\draw[->, thick, draw=black!60] (hitl) -- (browser);
\draw[->, thick, draw=black!60] (browser) -- (action);

\end{tikzpicture}
\caption{System architecture demonstrating the separation between AI compilation and zero-inference execution.}
\label{fig:architecture_improved}
\end{figure}
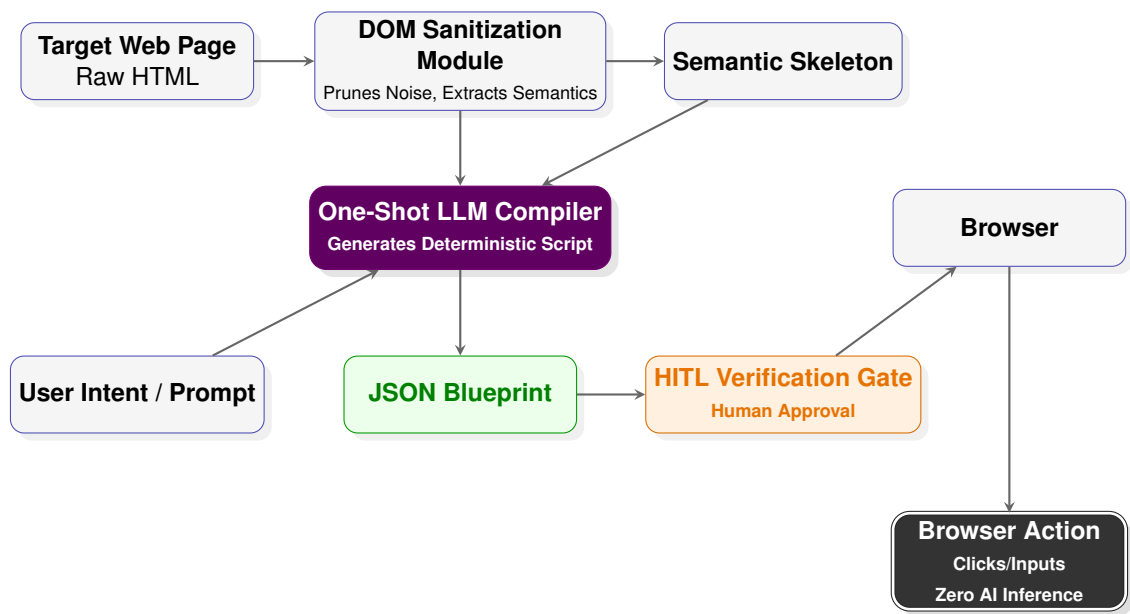

\subsection{Context Optimization via the DOM Sanitization Module (DSM)}

Supplying raw HTML to an LLM wastes tokens and degrades the model's attention mechanism via semantic noise. The DOM Sanitization Module (DSM) resolves this by acting as a context compression and signal extraction engine. Through a single DOM traversal, it applies three transformative operations:

\begin{itemize}
    \item \textbf{Noise Eradication:} Non-content subtrees (\texttt{<script>}, \texttt{<style>}, \texttt{<svg>}, and base64 payloads) are unconditionally pruned.
    \item \textbf{Signal Extraction:} Nodes styled as \texttt{display:none} or \texttt{visibility:hidden} are removed. Isolating visible business logic from hidden elements prevents the LLM from hallucinating interactions with non-interactive components.
    \item \textbf{Attribute Cleansing:} Volatile utility CSS classes are aggressively stripped, while semantic identifiers (e.g., BEM classes, \texttt{data-*} attributes) are preserved. This forces the LLM to ground its selectors in the application's permanent semantic structure, insulating the blueprint against UI redesigns and A/B tests.
\end{itemize}

Collectively, these operations compress token payloads by up to 85\%, ensuring target pages fit within the optimal reasoning windows of frontier models.

\subsection{One-Shot LLM Compilation and Selector Priority}

The sanitized HTML skeleton, active page URL, and user intent are submitted to the LLM in a single inference call. The system prompt strictly constrains the model to output a complete, executable JSON workflow schema without deferring decisions or requesting additional information. 

The prompt enforces a Semantic Selector Priority Hierarchy. The model must prefer robust semantic selectors (ARIA roles, \texttt{data-*} attributes, stable classes) over fragile positional paths like \texttt{nth-child}. This preference is operationally critical, as semantic selectors are substantially more resilient to front-end framework updates and A/B test variations. Within this single call, the LLM must also identify pagination patterns, construct loops, and emit extraction mappings. 

\subsection{Human-in-the-Loop Verification and Execution}

Prior to execution, the compiled JSON blueprint is presented to a human operator for review. This HITL gate serves as a safeguard against malformed or unintended action sequences, particularly those that might trigger irreversible side effects such as form submissions or account modifications. The operator may accept, reject, or manually amend the blueprint before authorizing execution. 

Upon approval, the lightweight execution engine traverses the JSON blueprint and issues the corresponding native browser API calls. The execution phase requires zero additional LLM inference calls; the connection to the model is fully terminated after compilation. To accommodate the asynchronous rendering behavior characteristic of Single Page Applications (SPAs), the engine employs dynamic wait heuristics---monitoring DOM mutation events and network-idle signals rather than fixed-duration sleep timers---improving reliability across dynamically rendered interfaces. 

This HITL boundary also serves as a deterministic manual fallback during execution halts. The blueprint's modular JSON structure allows for immediate manual patching either by directly injecting a corrected DOM selector or by utilizing a localized interaction recorder to bridge the point of failure without requiring a full system restart. 

\subsection{Lazy Replanning Architecture and Selector Healing}

To bridge the gap between deterministic execution and agentic resilience, the proposed architecture is designed to support a Lazy Replanning Architecture. Rather than functioning as a continuous agent, the system employs sparse LLM usage. When the execution engine encounters a terminal state---such as a missing DOM element resulting from a UI mutation---it does not fail silently. Instead, the deterministic failure serves as a trigger to invoke the LLM for targeted selector healing. 

Specifically, the execution engine halts and triggers the lazy replanning compiler under three distinct failure modes: 
\begin{itemize}
    \item \textbf{When the UI Changes:} The target application undergoes a structural mutation (e.g., an A/B test or framework update), causing the blueprint's semantic selector $l_i$ to resolve to null. 
    \item \textbf{When Execution Breaks:} Network latency, asynchronous SPA rendering timeouts, or unexpected modal popups interrupt the deterministic sequence. 
    \item \textbf{When the Plan Fails:} The blueprint successfully executes, but the extracted payload violates expected schema constraints, indicating a semantic misalignment during the initial compilation. 
\end{itemize}

The engine captures the mutated DOM state and routes it back to the LLM compiler for targeted selector healing. Crucially, this fallback mechanism does not revert the architecture to a continuous-loop agent. In continuous paradigms, the LLM dynamically governs the control flow of the workflow, evaluating state to determine the next logical action. In our Lazy Replanning architecture, control flow remains strictly encapsulated within the deterministic runtime. The LLM is invoked exclusively as an exception handler---resolving a null pointer (the invalidated selector)---without altering the compiled sequence of operations. 

While longitudinal empirical studies are required to mathematically prove the rarity of recompilation across the broader internet, isolating LLM invocation to exception handling guarantees that inference costs are strictly a function of structural UI volatility ($O(R)$), rather than the execution loop itself ($O(M \times N)$). This confines expensive LLM spatial reasoning strictly to moments of structural unpredictability.

\section{Economic Evaluation and Cost Benchmarks}

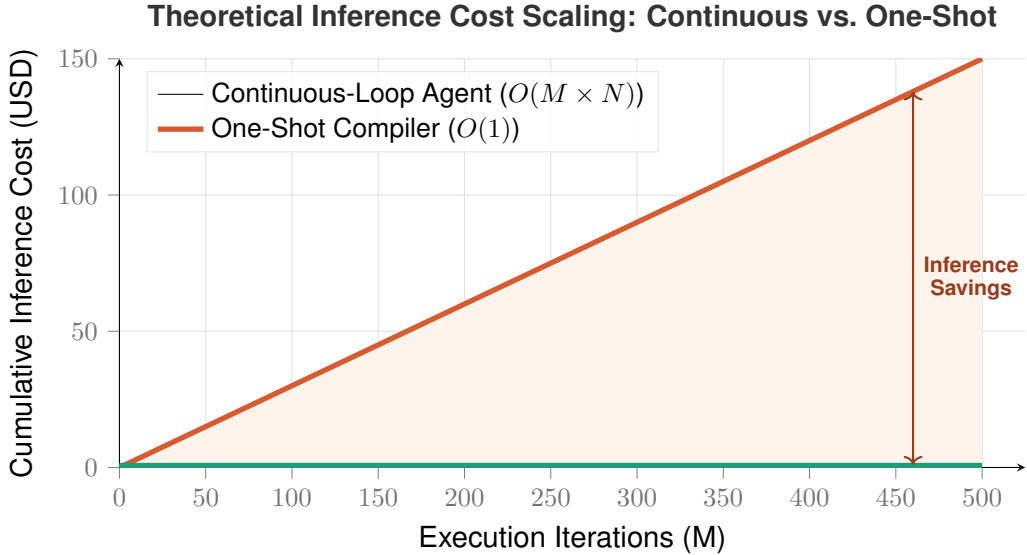
\begin{figure}[H]
    \centering
    \begin{tikzpicture}
      \begin{axis}[
        width=0.9\textwidth, height=7cm,
        title={Theoretical Inference Cost Scaling: Continuous vs.\ One-Shot},
        title style={font=\bfseries\sffamily, color=black!80},
        xlabel={Execution Iterations (M)},
        ylabel={Cumulative Inference Cost (USD)},
        xmin=0, xmax=525,
        ymin=0, ymax=150,
        grid=both,
        grid style={line width=.1pt, draw=gray!12},
        major grid style={line width=.2pt, draw=gray!25},
        legend pos=north west,
        legend style={
          cells={anchor=west}, font=\small\sffamily,
          draw=gray!20, fill=white, fill opacity=0.85, text opacity=1
        },
        axis lines=left,
        enlargelimits=false,
        tick align=outside,
        tick label style={font=\small\sffamily, color=black!55}
      ]

      \addplot[
        fill=red!10!orange!8, draw=none, domain=0:500
      ] {0.3 * x} \closedcycle;

      \addplot[
        domain=0:500, samples=2,
        color={rgb,255:red,216;green,90;blue,48},   
        line width=2pt
      ] {0.3 * x};
      \addlegendentry{Continuous-Loop Agent ($O(M \times N)$)}

      \addplot[
        domain=0:500, samples=2,
        color={rgb,255:red,29;green,158;blue,117},  
        line width=2.5pt
      ] {0.5};
      \addlegendentry{One-Shot Compiler ($O(1)$)}

      \draw[<->, thick,
  color={rgb,255:red,153;green,60;blue,29}
] (axis cs:460, 1) -- (axis cs:460, 138)
  node[midway, right, font=\scriptsize\bfseries\sffamily,
       text={rgb,255:red,153;green,60;blue,29}] {\shortstack{Inference\\Savings}};

      \end{axis}
    \end{tikzpicture}
    \caption{Cost comparison demonstrating linear scaling versus amortized constant cost.}
    \label{fig:cost_scaling}
\end{figure}

\subsection{Theoretical Cost Framework}

Let $N$ be the number of sequential actions in a single workflow, $M$ the number of execution iterations, $C_t$ the average LLM inference cost per token, and $S_i$ the token size of the DOM state at step $i$. For continuous-loop agents, the model evaluates the full state at every step of every execution: 

\begin{equation}
\text{Cost}_{cont} = M \times \sum_{i=1}^{N} [S_i \times C_t]
\end{equation}

This expression grows linearly in $M$: doubling the number of reruns doubles the total cost, independently of any per-step optimization. For the proposed compilation architecture, the model is invoked exactly once, over a sanitized skeleton of size $S_{compile}$: 

\begin{equation}
\text{Cost}_{oneshot} = 1 \times (S_{compile} \times C_t) + C_{exec}
\end{equation}

The baseline execution overhead $C_{exec}$ comprises standard compute and browser API calls; its monetary cost is negligible relative to frontier API fees. The one-shot architecture therefore achieves an effectively amortized $O(1)$ inference cost with respect to $M$: regardless of how many times the compiled blueprint is executed, no additional inference charges are incurred. 

\subsection{Applied Empirical Benchmark}

To ground the theoretical analysis, we instantiate the cost framework using a representative enterprise extraction task: five data fields per profile across 500 target profiles, assuming a raw DOM representation of 20,000 tokens per page. 

\begin{itemize}
    \item \textbf{Unoptimized continuous baseline:} Full DOM evaluation at each of five steps per profile yields approximately 2,500 sequential API calls and an estimated total cost of \$150.00. 
    \item \textbf{Optimized continuous baseline (90\% caching):} With aggressive DOM-diffing and state caching, inference calls are reduced but not eliminated, resulting in an estimated cost of \$15.00. 
    \item \textbf{One-shot compilation:} A single LLM invocation over the sanitized DOM skeleton yields a compiled blueprint; all 500 extraction iterations are executed via the deterministic runtime. Total inference cost ranges from \$0.002 to \$0.10 depending on model selection. 
\end{itemize}

This represents a cost reduction of up to 1500$\times$ compared to un-optimized continuous agents under equivalent execution workloads. The per-compilation costs observed across five frontier models are reported in Table \ref{tab:models}. 

\begin{table}[H]
    \centering
    \caption{One-shot compilation performance across leading LLMs (OpenRouter).}
    \label{tab:models}
    \renewcommand{\arraystretch}{1.2}
    \begin{tabularx}{\textwidth}{@{} >{\raggedright\arraybackslash}X c c c c @{}}
        \toprule
        \textbf{Model} & \textbf{Input $\to$ Output Tokens} & \textbf{Inference Cost (USD)} & \textbf{Speed (TPS)} & \textbf{Result} \\
        \midrule
        Claude Opus 4.6 & 11,628 $\to$ 1,340 & \$0.0916 & 96.9 & Success \\
        Claude Sonnet 4.5 & 11,628 $\to$ 1,670 & \$0.0599 & 98.6 & Success \\
        GPT-5.2-Codex & 9,951 $\to$ 1,447 & \$0.0377 & 115.7 & Success \\
        Qwen3.5 397B & 10,738 $\to$ 3,000 & \$0.0172 & 56.2 & Success \\
        Qwen3 Coder Next & 10,536 $\to$ 550 & \$0.0020 & 131.6 & Success \\
        \bottomrule
    \end{tabularx}
    
    \vspace{0.2cm}
    \small \textit{Note: All models listed successfully generated syntactically valid and executable JSON blueprints. Inference costs reflect a single compilation call over a sanitized DOM skeleton of approximately 10,000--12,000 tokens.}
\end{table}

Notably, ultra-low-cost models---specifically Gemini 2.5 Flash and Claude 3.5 Haiku---exhibited elevated failure rates in preliminary trials, attributable to insufficient reasoning depth in schema-constrained generation tasks. These models are excluded from Table \ref{tab:models} on the basis of systematic compilation failure rather than cost alone, a distinction relevant to model selection guidance in Section 5.

\subsection{Task Execution Evaluation}

To assess runtime reliability across representative enterprise automation scenarios, we evaluated the compiled execution engine on three distinct task modalities. For the extraction and fingerprinting modalities, 50 independent compilation attempts were performed; the form filling modality was evaluated across 10 complex configurations. Successful blueprints were executed against live target interfaces. 

Performance was assessed along two dimensions: compilation success rate (proportion of attempts yielding a syntactically valid, executable JSON schema) and execution accuracy (proportion of successful executions achieving the correct final browser state or data payload). 

\begin{itemize}
    \item \textbf{Task 1 --- High-Volume Paginated Extraction:} The system extracted 30 structured business profiles per page across 10 pagination states, targeting five fields per record (name, URL, address, website, and phone number). To simulate human interaction patterns and mitigate rate-limiting, the blueprint was executed with a 0.1-second stochastic inter-step delay and a 7-second inter-page delay. Compilation succeeded on 46 of 50 attempts; of these, 98\% achieved correct field extraction without hallucinated values. 
    \item \textbf{Task 2 --- Form Filling:} The system was evaluated on its capacity to map structured semantic payloads to obfuscated form fields, including non-standard input types and dropdown menus. Configurations requiring dynamic payload resolution via external webhooks completed within one minute. Compilation succeeded on 8 of 10 attempts, with 95\% execution accuracy across successful blueprints. 
    \item \textbf{Task 3 --- Technology Stack Fingerprinting:} The system navigated to target domains and analyzed sanitized DOM skeletons to identify deployed technologies, including CMS platforms, analytics trackers, and frontend frameworks, without relying on pattern matching heuristics. Compilation succeeded on 47 of 50 attempts, with 96\% execution accuracy. 
\end{itemize}

Across all modalities, compilation failures (ranging from 6\% to 20\%) were analyzed and taxonomized into three primary failure modes: (1) Schema Violations, where models emitted syntactically invalid JSON; (2) Semantic Misalignments, where the LLM selected a visually prominent but non-actionable DOM node; and (3) Reasoning Depth Exhaustion, specifically in complex form-filling (Task 2), where the model failed to map multi-step conditional dependencies. This taxonomy confirms that compilation failures are fundamentally bounded by the spatial reasoning limits of the underlying LLM, rather than flaws in the deterministic execution engine. 

Crucially, a zero-shot compilation failure does not equate to a system failure. Because the compiler emits a declarative JSON blueprint rather than an opaque continuous execution trajectory, errors are highly localized and human-readable. The Human-in-the-Loop (HITL) gate allows operators to manually patch isolated failures---such as correcting a single misaligned semantic selector---in seconds. This collaborative paradigm shifts the evaluation metric from pure zero-shot autonomy to overall system usability, enabling near-100\% execution reliability while completely preserving the amortized $O(1)$ inference cost.

\begin{table}[H]
    \centering
    \caption{Internal evaluation results across three enterprise task modalities.}
    \label{tab:results}
    \renewcommand{\arraystretch}{1.2}
    \begin{tabularx}{\textwidth}{@{} >{\raggedright\arraybackslash}p{5.5cm} >{\centering\arraybackslash}X >{\centering\arraybackslash}X >{\centering\arraybackslash}X @{}}
        \toprule
        \textbf{Task Modality} & \textbf{Compilation Attempts} & \textbf{Successful Blueprints} & \textbf{Execution Accuracy} \\
        \midrule
        T1: High-Volume Extraction & 50 & 46 (92\%) & 98.0\% \\
        T2: Form Filling & 10 & 8 (80\%) & 95.0\% \\
        T3: Technology Stack Detection & 50 & 47 (94\%) & 96.0\% \\
        \bottomrule
    \end{tabularx}
    
    \vspace{0.2cm}
    \small \textit{Note: Successful Blueprints indicates the count of compilations yielding a valid, executable JSON schema. Execution Accuracy indicates the proportion of those executions achieving the correct payload or final browser state.}
\end{table}

\section{Discussion and Limitations}

\subsection{Scope of Empirical Validation}

While the reliance on proprietary enterprise DOMs limits independent replication of the specific execution payloads, this constraint represents a deliberate methodological choice rather than an experimental oversight. Standardized macro-benchmarks such as WebArena or Mind2Web are designed to evaluate the generalized, zero-shot reasoning capabilities of agents navigating highly diverse and novel environments. Recent works such as Agent Workflow Memory (Zheng et al., 2024) and Explorer (Pahuja et al., 2025) heavily rely on these benchmarks to evaluate exploratory agent capabilities in dynamic, unstructured settings. 

However, the Compile-and-Execute architecture does not aim to advance zero-shot reasoning; rather, it addresses the economic scaling constraints of highly repetitive, structured enterprise workflows. Evaluating this architectural claim directly against benchmarks such as Mind2Web would introduce confounding factors between the underlying LLM’s spatial reasoning limitations and the architecture’s fundamental cost-scaling properties. By isolating the evaluation to a targeted micro-benchmark of representative enterprise tasks, we decouple the amortized $O(1)$ economic analysis from the variability introduced by generalized agent benchmarks. This evaluation therefore establishes a controlled empirical validation of the architecture’s amortized constant cost execution model under deterministic workflow conditions.

Future work should evaluate the cost-versus-reliability trade-off in volatile UI environments using macro-benchmarks. However, establishing baseline deterministic execution bounds requires the controlled enterprise setting utilized in this study.

\subsection{Cost-Resilience Trade-off}

While continuous-loop agents can theoretically recover from UI mutations in real time, the practical risk of deterministic blueprint failure in enterprise environments is frequently overstated. Structural UI mutations such as framework migrations or complete layout redesigns occur on macro-timescales, typically bi-annually or annually. Conversely, transient UI volatility, such as A/B testing or cosmetic CSS updates, rarely alters the underlying semantic HTML structure. Because the DOM Sanitization Module (DSM) explicitly grounds the compiled blueprint in stable semantic identifiers (e.g., ARIA roles and \texttt{data-*} attributes), the blueprint is inherently insulated against routine cosmetic deployments. Consequently, the frequency of mandatory recompilation is operationally negligible, ensuring the amortized $O(1)$ inference cost remains fully intact in practical enterprise deployments. 

Deterministic failure is, however, operationally preferable to probabilistic hallucination in enterprise contexts. Continuous agents naturally suffer from context window exhaustion over extended iterations; as they accumulate trajectory history, they begin to "forget" original constraints and become highly prone to logic drift. A continuous agent that encounters an unexpected UI state may continue executing with degraded confidence, potentially performing unintended or destructive actions. Conversely, a compiled blueprint executes with 100\% mechanical fidelity. Furthermore, the proposed architecture enforces strict timeouts and halts cleanly on unresolved selectors. Rather than a terminal flaw, this predictable halting is utilized as the precise trigger for the hybrid fallback and selector healing mechanisms described in Section 3.4. 

\subsection{Applicability Constraints}

The DOM Sanitization Module relies on the presence of structured semantic HTML to extract a meaningful skeleton. This assumption is violated by canvas-rendered applications, WebGL environments, and other viewport-driven interfaces that do not expose their interactive structure through the DOM. For such interfaces, multimodal continuous agents---such as the massive exploration-driven synthesis approach demonstrated by Explorer (Pahuja et al., 2025)---retain a distinct advantage, as they can operate over pixel-level visual representations rather than DOM semantics. The proposed architecture is therefore most applicable to conventional server-rendered or SPA-based web applications with standard HTML structure. Agentic Compilation should thus be viewed as a specialized, low-cost compiler for standard HTML workflows, deeply complementary to multimodal exploration frameworks used in unstructured environments.

\subsection{Abstraction of Technical Intermediaries}

A secondary but consequential advantage of the Compile-and-Execute paradigm is the abstraction of technical maintenance and deployment friction. Traditional RPA and LLM generated imperative scripts impose a high total cost of ownership, as they require specialized engineering personnel for workflow creation, infrastructure scaling, and ongoing maintenance. The proposed architecture substantially reduces this technical dependency across three layers: 

\begin{itemize}
    \item \textbf{Workflow Authorship:} By delegating the technical complexities of DOM navigation such as selector engineering, pagination inference, and loop deduction entirely to the one-shot LLM compiler, the architecture allows domain experts (e.g., operations or finance personnel) to author complex automations using natural language intent.
    \item \textbf{Workflow Maintenance:} In traditional automated pipelines, structural UI mutations necessitate code-level intervention from engineering teams. Under the proposed framework, the deterministic halting and hybrid fallback mechanisms (Section 3.4), coupled with HITL manual patching capabilities (Section 3.3), provide a localized, code-free recovery path for end-users.
    \item \textbf{Execution Scaling:} Because the compiled, amortized $O(1)$ JSON blueprint is evaluated by a lightweight runtime engine, scaling an automation from a single execution to hundreds of iterations requires no underlying server infrastructure provisioning or script optimization.
\end{itemize}

This democratization of automation is presently constrained to the approximately 80\% of enterprise workflows characterized by stable, standard HTML structures. Highly dynamic, canvas-rendered, or aggressively obfuscated interfaces remain outside the scope of code-free abstraction and continue to require specialized technical judgment.

\begin{figure}[H]
    \centering
    \begin{tikzpicture}[node distance=1.5cm and 1cm, >=stealth, scale=0.85, transform shape]
      \tikzstyle{comp} = [rectangle, rounded corners=0.15cm, draw=violet!50, fill=violet!5, thick, minimum width=3.5cm, minimum height=1.2cm, align=center, text=violet!90!black, font=\sffamily, drop shadow={opacity=0.08}]
      \tikzstyle{exec} = [rectangle, rounded corners=0.15cm, draw=teal!50, fill=teal!5, thick, minimum width=3.5cm, minimum height=1.2cm, align=center, text=teal!90!black, font=\sffamily, drop shadow={opacity=0.08}]
      \tikzstyle{decision} = [diamond, aspect=1.8, rounded corners=0.1cm, draw=orange!60, fill=orange!10, thick, text=orange!90!black, align=center, font=\bfseries\sffamily, inner sep=1pt, drop shadow={opacity=0.08}]
      \tikzstyle{succ} = [rectangle, rounded corners=0.4cm, draw=teal!70!black, fill=teal!60!black, thick, text=white, minimum width=2.2cm, minimum height=0.8cm, align=center, font=\bfseries\sffamily, drop shadow={opacity=0.15}]
      \tikzstyle{fail} = [rectangle, rounded corners=0.15cm, draw=red!50, fill=red!5, thick, minimum width=3.5cm, minimum height=1cm, align=center, text=red!80!black, font=\sffamily, drop shadow={opacity=0.08}]
      \tikzstyle{recomp} = [rectangle, rounded corners=0.15cm, draw=violet!70!black, fill=violet!60!black, thick, text=white, minimum width=3.2cm, minimum height=1cm, align=center, font=\bfseries\sffamily, drop shadow={opacity=0.15}]

      \node[comp] (comp) at (0,0) {\textbf{Compilation}\\LLM $\to$ JSON Blueprint};
      \node[exec] (exec) at (5,0) {\textbf{Deterministic Execution}\\Execute Actions in Browser};
      \node[decision] (dec) at (10,0) {UI Change /\\Error?};
      \node[succ] (succ) at (13.5,0) {Success};

      \node[fail] (fail) at (10,-2.5) {\textbf{Failure Handling}\\UI Update Detected};
      \node[recomp] (rebuild) at (7,-4.5) {Rebuild JSON\\Blueprint};

      \draw[->, thick, draw=gray!60] (comp) -- (exec);
      \draw[->, thick, draw=gray!60] (exec) -- (dec);
      \draw[->, thick, draw=gray!60] (dec) -- node[above, font=\scriptsize\sffamily\bfseries, text=gray!80!black] {No} (succ);
      \draw[->, thick, draw=gray!60] (dec) -- node[right, font=\scriptsize\sffamily\bfseries, text=gray!80!black] {Yes} (fail);
      \draw[->, thick, draw=gray!60] (fail) -- node[above left, font=\scriptsize\sffamily, text=gray!80!black] {Recompile} (rebuild);
      
      \draw[->, thick, draw=gray!60] (rebuild.west) -| (comp.south);
      
    \end{tikzpicture}
    \caption{Lazy Replanning Architecture and UI Fallback Loop.}
    \label{fig:fallback}
\end{figure}
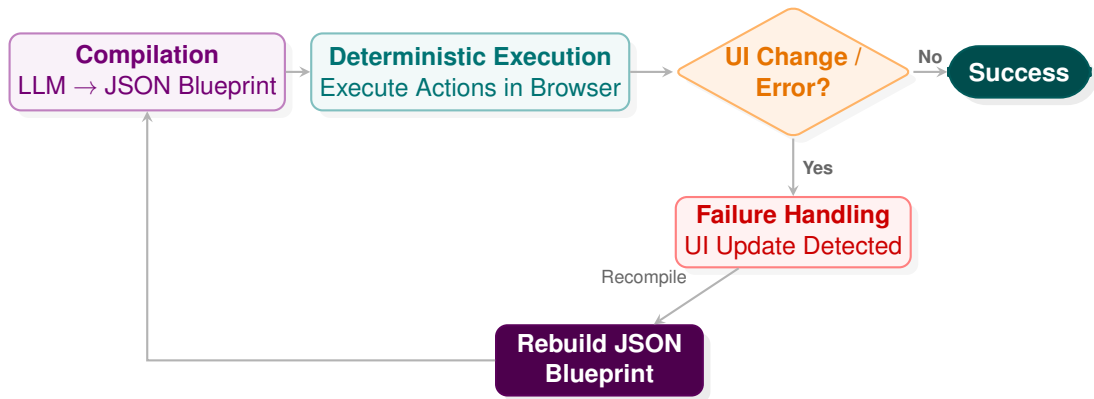

\subsection{Future Directions}

Several extensions of the current framework merit investigation. First, automated recompilation---triggering a new compilation call when execution halts on a missing selector---would partially recover the resilience advantages of continuous agents while preserving the amortized $O(1)$ cost property for stable interfaces. Second, the integration of lightweight DOM change detection could signal when recompilation is warranted, reducing unnecessary recompilation overhead. Third, formal evaluation against established web agent benchmarks would enable rigorous comparison with prior work and support broader adoption of the proposed paradigm. 

\section{Conclusion}

We have presented a Compile-and-Execute architecture for web automation that resolves a fundamental economic inefficiency in contemporary continuous-loop agent deployments. By confining LLM involvement to a single compilation phase---wherein user intent and a sanitized DOM representation are translated into a deterministic JSON workflow blueprint---the architecture reduces inference cost scaling from $O(M \times N)$ to amortized $O(1)$ with respect to execution iterations. 

Empirical evaluation across five frontier models and three enterprise task modalities demonstrates zero-shot compilation success rates of 80--94\%. Because the resulting blueprints are modular and directly editable, minimal Human-in-the-Loop (HITL) intervention elevates this to near-100\% operational reliability prior to execution, at per-compilation costs between \$0.002 and \$0.092. Crucially, the architecture is model-agnostic at the compilation layer. The $O(1)$ cost guarantees are architectural and permanent, while operational accuracy will naturally scale with future improvements in baseline model capability. The architecture does not claim to supersede continuous-loop agents in all settings; for tasks requiring real-time adaptability to volatile or visually rendered interfaces, such agents remain preferable. Rather, the contribution is a principled identification of the class of tasks---stable, repetitive, structure-amenable workflows---for which deterministic compilation is both sufficient and substantially more efficient. 

The proposed paradigm extends beyond web automation to any deterministic, repeatable workflow where planning can be decoupled from execution. By making this distinction explicit and providing a concrete implementation of the compilation paradigm, this work constitutes a meaningful step toward large-scale, economically viable automation at scales previously infeasible under continuous inference architectures.

\end{document}